This is the pre-peer reviewed version of the following article

# Attention-driven read-aloud technology increases reading comprehension in children with reading disabilities

Gianluca Schiavo[1], Nadia Mana[1], Ornella Mich[1], Massimo Zancanaro[1,2] and Remo Job[2]

[1]FBK - Fondazione Bruno Kessler, Via Sommarive 18, Trento, Italy

[2]University of Trento, Department of Psychology and Cognitive Science, Corso Bettini, Rovereto, Italy

## Abstract

The paper presents the design of an assistive reading tool that integrates read-aloud technology with eye-tracking to regulate the speed of reading and support struggling readers in following the text while listening to it. The paper describes the design rationale of this approach, following the theory of auditory-visual integration, in terms of an automatic self-adaptable technique based on the reader's gaze that provides an individualized interaction experience.

This tool has been assessed in a controlled experiment with 20 children (aged 8-10 years) with a diagnosis of dyslexia and a control group of 20 children with typical reading abilities. The results show that children with reading difficulties improved their comprehension scores by 24% measured on a standardized instrument for the assessment of reading comprehension, and that children with more inaccurate reading (N=9) tended to benefit more.

The findings are discussed in terms of a better integration between audio and visual text information, paving the way to improve standard read-aloud technology with gaze-contingency and self-adaptable techniques to personalize the reading experience.

*Keywords*: human-computer interface, special needs education, reading difficulties, dyslexia, eye tracking, assistive technology

## Introduction

The term *developmental reading difficulties* is used to describe a variety of obstacles that affect the ability to read in children. These obstacles might include difficulties with accurate and fluent word recognition, as well as poor spelling and decoding abilities or poor reading comprehension. In this work, we focused on the specific learning disability called *developmental dyslexia*, which is a learning difficulty that primarily affects the skills involved in accurate and fluent word reading and spelling and it is characterized by difficulties in phonological awareness, verbal memory and verbal processing speed (Rose, 2009).

Developmental dyslexia is a neurodevelopmental disorder with genetic, neurobiological, and cognitive components (Hulme and Snowling, 2016) and with a prevalence of more than 10% of the population (Gabrieli, 2013; Franceschini et al., 2015). Dyslexia is characterized by, but not limited to, difficulties with decoding written input. Specifically, people with dyslexia present problems in associating written letters (graphemes) with their specific sounds (phonemes) and in relating the sounds of language to letters and words, a mapping that may be quite complex even in shallow orthographies (Job, Peressotti, & Cusinato, 1998, Tilanus et al., 2016). These difficulties lead to slow and error-prone reading, misspelled words and difficulties in identifying and remembering complex, uncommon and new words as well as fatigue and stress after reading for a short time which eventually have an impact on the reading performance and reading comprehension (Kim, Linan-Thompson, & Misquitta, 2012; Snow, 2002). According to the Simple View of Reading (Gough & Tunmer, 1986), reading comprehension skill is the product of decoding and listening comprehension. Comprehension difficulties in dyslexia are indeed consequential to the input decoding difficulties (Peterson & Pennington, 2012): some children with dyslexia have problems with reading comprehension, which are attributable to slow and inaccurate word reading, leaving few attentional resources available for comprehension (Hulme & Snowling, 2016, Snowling, 2013). Subtypes of dyslexia can be identified according to the specific orthography-to-phonology mapping involved (Friedmann & Coltheart, 2016). If not properly identified and addressed, reading difficulties continue to affect throughout adulthood with impact in learning and workplace activities and resulting in a lack of self-confidence and self-worth (Shaywitz, 2008).

Quite a lot of experimental and clinical work has been done in order to establish whether the underlying causes of dyslexia are linguistic, e.g. a phonological deficit (Liberman & Shankweiler, 1985; Ramus & Szenkovits, 2008; Vellutino et al., 2004; Ziegler & Goswami, 2005; Castles, Coltheart, Wilson, Valpied, & Wedgwood, 2009), or rather emerge from dysfunctional cognitive processes, such as perceptual/attentional problems (Bosse, Tainturier, & Valdois, 2007; Romani, Tsouknida, di Betta, & Olson, 2011; Ruffino, Gori, Boccardi, Molteni, & Facoetti, 2014), working memory and executive impairments (Brosnan et al., 2002; Kibby, Marks, Morgan, & Long, 2004; Smith-Spark & Fisk, 2007), implicit learning difficulties (Vicari et al., 2005; Pavlidou, Kelly, & Williams, 2010), or serial order processing (Szmalec, Loncke, Page, & Duyck, 2011; Hachmann et al., 2014).

Reading disabilities and attentional disorders often co-occur (Willcutt, 2018). Some recent works suggest a link between reading difficulties and attention in terms of a multiple deficit model of developmental disorders (Pennington, 2012; Willcutt, 2018; Langer et al. 2019) that have strong implications on intervention approaches (Ring & Black, 2018). Along this line, a recent review work by Vidyasagar (2019) identified in visual attention an important aspect for developing effective remediation strategies for addressing developmental dyslexia, suggesting that interventions training visuo-spatial attention and temporal sampling might be effective in improving reading performance. In parallel with this aspect, it is important to acknowledge that readers with dyslexia present a generally good oral comprehension (Kida et al., 2016) that can be exploited in intervention. Although there is still debate on the causes of dyslexia, practitioners agree on considering interventions based on phonetics, reading fluency, and phonemic awareness training as effective ways to lessen the reading difficulties (Magnan & Ecalle, 2006; Galuschka, Ise, Krick, & Schulte-Körne, 2014; Hulme & Snowling, 2016), even considering technology-based training intervention (see for example, Van de Ven et al., 2017; Hautala et al., 2020; Ronimus et al., 2020). Yet, the difficulties met during reading often discourage children from practising and led to consequent reluctance to read (Justice, 2006; Muter & Snowling, 2009).

Among the strategies that support readers with dyslexia, technology-based reading interventions have been proven to be beneficial for struggling readers. Indeed, recent meta-analyses suggest that students with reading difficulties benefit when their reading is supported by reading-aloud tools (Buzick & Stone, 2014; Li, 2014; Wood, Moxley, Tighe, & Wagner, 2018). The effect size for students with learning disabilities is larger than the effect size for students without disabilities and it is even larger when the text is read by a human voice rather than by an automatic voice based on speech synthesis - Human voice: estimated effect sizes .61 (with learning disability) vs .48 (without learning disability); Speech synthesis: .39 (with learning disability) vs .26 (without learning disability) (Li, 2014). Furthermore, a study by Grunér and colleagues investigates the compensatory effect of TTS technology on reading comprehension in children with reading disabilities, with or without co-occurring problems with inattention and hyperactivity, showing that symptoms of inattention and hyperactivity can moderate the effect of text-to-speech technology on reading comprehension. (Grunér, Östberg, & Hedenius, 2018). In a recent review by Jamshidifarsania and colleagues (2019), it has been argued that the potentiality of intelligent self-adaptable systems to assess the student's skills and provide individualized interaction modalities has not yet been adequately explored.

One of the limits of existing tools is that they fail to keep track of the portion of text visually processed by the reader such that there may be an incongruence between what the reader reads and what he/she hears. Yet, recent works (see, for example, Valentini et al. 2018; Knoop-van Campen Segers, & Verhoeven, 2020) suggest that simultaneously listening to and reading stories leads to better comprehension than using a single modality. This might be beneficial also for readers with dyslexia: presenting information in two modalities might benefit working memory by providing a redundant but compatible input so that more attention can be paid to comprehension and encoding of word meanings. This is also supported by the theory of auditory–visual integration (Goswami, 2011; Schneps et al., 2019), suggesting that accurate and rapid word decoding builds upon successful content and time integration between auditory and visual information. In this respect, people with reading impairments can make effective use of concurrent visual and auditory stimuli, especially when using assistive tools such as read-aloud technology.

Our research question focuses on the possibility of using gaze position as a proxy for attention in a reading task to automatically synchronize the reading aloud of the text to the actual reading of the text and to support the integration of visual and auditory information. We believe that this adaptation enforces the bimodal experience of listening and reading and the integration between auditory and visual information which is known to be beneficial (Valentini et al. 2018, Schneps et al., 2019, Knoop-van Campen et al., 2020). In this paper, we present the rationale for the design of an assistive reading tool which integrates an eye-tracker to a standard read-aloud tool and its evaluation with a controlled study.

In the next section, we briefly summarize the state of the art on assistive technology for reading difficulties and present an assistive reading device named GARY (Gaze And Read it by Yourself) that integrates an eye tracker with a read-aloud tool to implement a self-adaptable system aimed at providing individualized interaction based on visual attention cues. Then, we present a controlled study to assess the short-term benefit of our approach by comparing the effects on text comprehension with respect to a standard read-aloud tool. Finally, we discuss the results of the study and some implications for the design of remediation approaches exploiting the role that attention plays in reading difficulties.

## Related works

Most individuals with reading difficulties are encouraged to use additional support to deal with their reading difficulties. Assistive technology, i.e. specific hardware and/or software tools designed to support people with disabilities, might help readers with dyslexia to access content and information in different ways, mitigating the challenges associated with reading, writing, and spelling (Nordström, Nilsson, Gustafson & Svensson, 2019; Dawson, Antonenko, Lane, & Zhu, 2019; Reid, Strnadová, & Cumming, 2013; Hasselbring, T. S., & Bausch, M. E., 2005). Common assistive technology for individuals with dyslexia, such as text-to-speech (TTS) and read-aloud tools, are used for addressing the problems with rapid word recognition and poor spelling and decoding abilities associated with dyslexia. These tools support people with reading difficulties by offering a text presentation, where the to-be-read text is paired either with synthesized computer voice (TTS tool) or audio recording of the same text read by humans (read-aloud tool).

Attention-driven Read-aloud Technology

Text-to-speech and read-aloud tools have spread since the 80s (Klatt, 1987; Crawford & Tindal, 2004; Biancarosa & Griffiths, 2012). However, as computer technology has significantly improved over the last twenty years, the number of tools of this type which are of good quality and easily accessible is significantly increasing (e.g., Kurzweil 3000, MWSReader). These tools commonly include features such as reading speed regulator, voice and pronunciation selector, creation of synthetic audio files, and dynamic text highlighting (Biancarosa & Griffiths, 2012), exploited by students with reading difficulties to support their reading experience. A recent meta-review (Wood et al., 2018) has found that the use of TTS and related read-aloud tools improves reading comprehension for students with reading disabilities, even though the underlying cognitive mechanisms remain largely unknown. Schneps and colleagues (2019) propose that theories of auditory-visual integration underlying dyslexia (Goswami, 2011) can explain the advantage of read-aloud tools that combine concurrent visual and auditory stimuli. According to this theory, accurate and rapid word decoding is achieved through successful content and time integration between auditory, visual and motor control, and dyslexia might stem from a deficit in such integration/synchronization process. TTS technology can contribute in supporting the multimodal integration by providing access to alternative input streams (visual and auditory) for language processing in instances where one input becomes weaker during the reading process (Schneps et al., 2019). Moreover, the ability of TTS software to visually highlight words coupled with the audio pronunciation might improve verbal and visual information processing in working memory, increasing the mental resources that can be devoted to comprehension.

Nevertheless, TTS technology suffers from inherent limitations. For example, TTS are driven to rates that are generally slower than the speed usually attained in typical mental reading. Moreover, TTS speed rate is usually fixed to a constant value that can be adjusted only manually by the users by interrupting the reading process. In this article, we investigate the effect of the combination of an attention-driven read-aloud technology with eye tracking on reading comprehension for young children with reading difficulties in order to allow the possibility for the reader to smoothly follow the text and to integrate both visual and auditory information streams.

Using eye tracking to track the flow of reading in real time has been a topic of much research. Eye tracking devices are used to localize the eyes position and to track their motion. Information of the eye movements coupled with the estimation of the head position allows to determine the gaze direction and to locate the point of the reader's gaze on a computer screen (Morimoto & Mimica, 2005). In the last decade, many eye-tracker applications have been developed for augmenting the reading experience (Biedert, Buscher, & Dengel, 2009), or for assisting users with motor disabilities or visual impairments (Duchowski, 2002). Nevertheless, the use of eye tracking for supporting struggling readers has not been extensively investigated and only a few applications have been developed for gaze-contingent adaptation of the text-presentation (Lunte, & Boll, 2020) and for screening and diagnostic purposes (Nilsson Benfatto et al., 2016). With respect to this latter aspect, our approach is different since information from the readers' gaze is used not for diagnostic or clinical goals but for supporting the digitally mediated reading process.

### GARY: Speech synthesis and eye tracking to support struggling readers

GARY is an assistive reading device that integrates a display for the written content, a loudspeaker, and an eye-tracker. It works as a read-aloud accommodation tool with the textual content displayed using proper fonts to facilitate reading. In particular, the text is displayed with 12-point Arial font with a line spacing of 1.5, following the indication of the Dyslexia Style

Guide (British Dyslexia Association, 2018). The system also allows the reading aloud of the textual content and each phrase (from 1 to 5 words, depending on the prosodic patterns of the text) is highlighted on the display while it is read. While the content is read aloud, the eye-tracker locates the user's point of gaze, monitoring if the user is looking at words following the highlighted phrase. If the reader gazes at the following portion of text, the reading continues to the next phrase. Otherwise, the reading stops. In order to deal with inaccuracies of eye tracking measurements, fixations are considered in an area that extends the underlying text size by a factor that takes into account the tracking error.

The key aspect of GARY is that the audio speed is automatically adjusted to keep up with the reading speed of the user, and therefore driven by the user's attention to the text.

Three main design principles guided the development of GARY. The first one is the interaction between the reading and listening processes. The assumption here is that if a struggling reader is able to keep pace with the reading aloud track while silently reading herself, she will (a) gain a better comprehension of the text than from a simpler audio content only and (b) support and improve her reading abilities. Evidence for this assumption are based on the auditory-visual integration theory and can be found in a recent meta-analysis suggesting than text-to-speech and read-aloud presentation positively affects reading comprehension for individuals with reading disabilities (Wood et al., 2018).

The second design principle refers to the control of attention. Differently from a traditional read-aloud software, GARY uses information from the position of the reader's gaze to guide the reading pace. In this respect, the specific implementation of GARY aims at maintaining the reader's visual attention on the portion of text while it is read aloud by interrupting the auditory reading aloud when the participant's eyes fixate away from the text actually read aloud by the speech-synthesis software. Of course, eye fixation might not fully coincide with covert attention but in the reading process it may be a reasonable evidence for it (Rayner, 2009). This functionality might lead the reader to focus on the text read by the TTS, promoting the concurrent integration between auditory and visual information, allowing to resolve any mismatch of the two processes in favour of the read-aloud text.

Finally, the third design principle relates to the coupling of the child's needs and the support provided. As a scaffolding artefact (Sharma & Hannafin, 2007), GARY adapts to the child's needs based on an ongoing assessment of visual attention. This principle is implemented through the gaze-regulated feature that dynamically adapts the reading aloud to the individual pace instead of adopting a constant predefined speed, like in traditional reading aloud and text-to-speech applications.

The current prototype (Figure 1) is realized with a tablet PC and an infrared eye-tracker encased in a wooden frame. Because of the current limitation of the specific low-cost eye-tracker used, the frame is positioned on a table rather than handheld. The tilting of the display and the eye-tracker can be adjusted. Furthermore, as in most eye-tracking applications, the eye-tracker needs to be individually calibrated before use in order to tune the system to the individual and environmental conditions. The calibration phase takes a few minutes and requires completing a task of looking at twelve dots as they appeared on the screen. Although these are undoubtedly limitations for a normal usage, they might be considered acceptable for the sake of an experimental condition such as the one presented in this study.

Following Li (2014), in order to improve the benefit of reading aloud, GARY does not use an automatic speech synthesis but a read-aloud component based on audio pre-recorded by a semi-professional voice artist and manually aligned with the text. As in common applications on the market, GARY includes a highlighting functionality to point out the words while they are read (Figure 1). Word highlight and voice playback occur in parallel and proceed when the participant's eyes fixate the next groups of words.

Before the empirical study described in the following section, in order to assure proper accuracy, reliability and robustness of the system, and to refine the study procedure, GARY was preliminary tested in local Primary schools with 40 pupils, 20 with and 20 without reading difficulties.

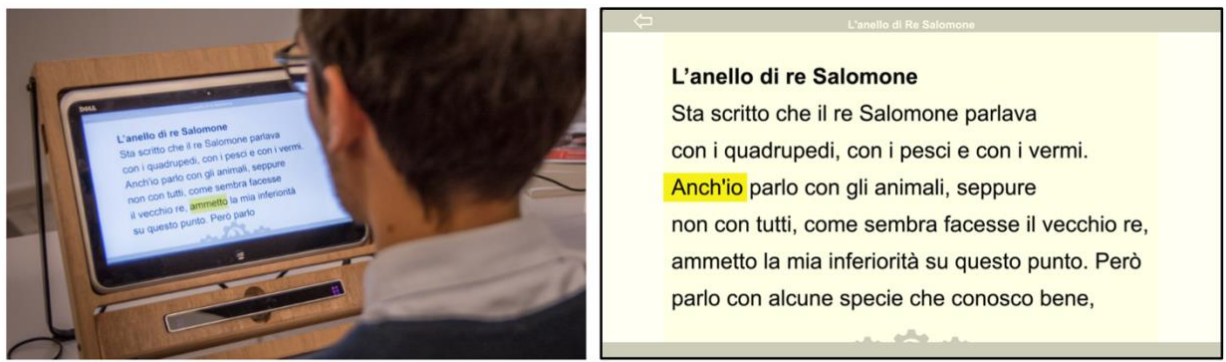

[Figure 1. GARY apparatus (left) and a detail of the display (right).]

## The Study

An empirical study was conducted to investigate the effect of controlling the reading aloud speed by monitoring the gaze of young children with dyslexia. The study tests the hypothesis that readers with dyslexia would benefit from using this type of read-aloud technology, in terms of reading comprehension, whereby no benefit is expected for typical readers. According to the auditory–visual integration, pairing the reading and listening processes should allow readers with dyslexia to keep track of the text, which, in turn, may avoid overtaxing working memory in terms of processing time, and allow to resolve any mismatch of the two processes in favour of the read-aloud text. This would, on the one hand, improve working memory efficiency and, on the other hand, prevent incorporating reading errors into the process, thus favouring a better comprehension of the text.

For typical readers, though, as their reading pace and accuracy are already quite good, no difference should be expected in terms of reading comprehension from the use of read-aloud technology.

The following hypothesis is therefore being tested:

*Hp: Readers with dyslexia have a better comprehension using GARY with respect to a traditional read-aloud tool.*

While typical readers are not expected to increase their comprehension using GARY with respect to a traditional read-aloud tool, a sample of typical readers have been collected as a baseline for assessing the expected improvement of dyslexic readers and to permit direct assessment of the efficacy of this approach specifically for children with dyslexia.

The hypothesis was tested with a 2×2 factorial mixed model design including *group* as between-subject variable (readers with dyslexia and typical readers) and *type of technology* as within-subject variable (gaze-regulated read-aloud tool vs. traditional read-aloud tool). *Reading comprehension scores* on a standardized test comprised the dependent variable. *Reading speed* (measured as syllables per second) was measured to control for potential effect on reading comprehension. A linear mixed model approach (Garson, 2013) was chosen to inspect the hypothesized relationship in a more adequate manner.

Moreover, we expect that GARY might provide more benefits to readers with more severe difficulties and less accurate reading performance. Readers with dyslexia with inaccurate reading are characterized by a weak decoding ability, thus the use of a gaze-regulated read-aloud tool might help to direct attention to the text and facilitate word recognition by, e.g., ruling out visually similar competitors, thus better supporting their reading comprehension compared to readers with milder reading difficulties. To better understand such performance variations within the experimental group, we subsequently analysed variations between the type of technology in terms of gain scores of reading comprehension and performance on the standardized diagnostic tests.

**Technology and apparatus**

Two different versions of the read-aloud tool were tested in the "type of technology" condition (Traditional and GARY).

In the Traditional condition, the software simply highlighted the text while it was read aloud by the system (without using the eye-tracking adaptation). Therefore, the tool worked as a traditional read-aloud application and usual read-aloud controls were provided for manually controlling the audio playback.

In the GARY condition, the read-aloud tool integrated the eye-tracking function and used gaze information for guiding the reading pace as explained above. In this condition the text is highlighted while it is read aloud by the recorded voice and the software paces through the text following the reader's gaze. The reading proceeds if the user looks at words following the highlighted text, otherwise the reading is paused (see Figure 2).

The same apparatus (i.e. the prototype described in Section 3) was used in both conditions: the only variation was that reading speed can be manually controlled with interface commands in the Traditional condition, whereas it is automatically guided by data from the gaze tracking in the GARY condition (as described above). In both conditions the words read by the application (1-5 words, according to a manual pre-segmentation of the text based on the prosodic

Attention-driven Read-aloud Technology

phrasing) were highlighted in yellow and the interface was kept as similar as possible: the text was displayed in the same way in both conditions as explained above.

**GARY condition**

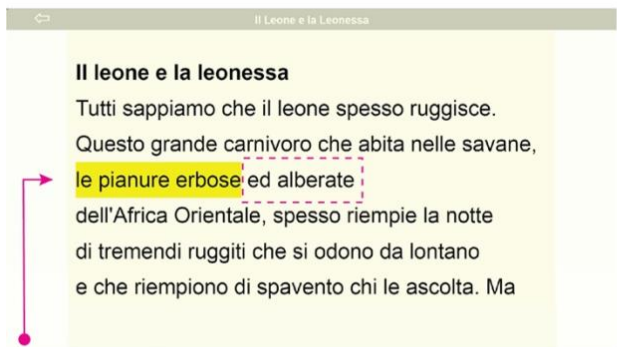

In the GARY condition, as the program moves through the text following the gaze of the reader, the text is highlighted and it is read-aloud by a recorded voice. The reading proceeds if the user looks at the words following the highlighted text (the area within the dashed square, not visible in the interface), otherwise the reading is paused.

**Traditional condition**

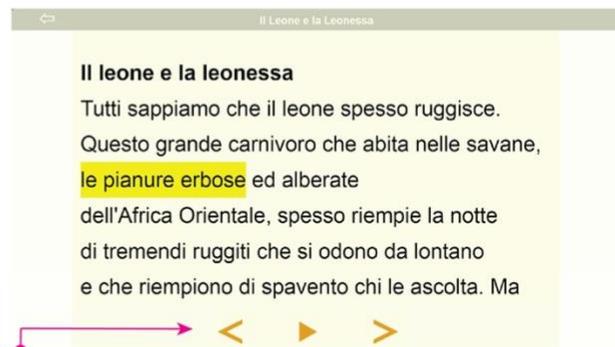

In the Traditional condition, the text is highlighted while it is read-aloud by a recorded voice. The software moves through the text at a predefined speed and the read-aloud voice can be controlled using the command on the bottom part of the interface  (play / pause / skip forward / skip backward).

[Figure 2. An example of a story page in the GARY (left) and Traditional (right) condition. The main difference in the interface are the on-screen commands included in the Traditional condition.]

**Participants**

The participants were 20 Italian children[1] (6 F / 14 M) with a history of reading struggles and a diagnosis of dyslexia (experimental group) and 20 Italian children (10 F / 10 M) with typical reading abilities (control group). All participants were primary school students from the same urban area, with an average age of 9.3 years (SD= 0.7). The two groups did not differ with respect to age (Mann-Whitney test: U=261, *p* >.05).

The participants in the experimental group were recruited in a clinical centre for students with learning disabilities that they regularly attend; the sessions took place in the same centre.

The children in the control group were recruited through personal contact in local schools and in after-school centers and the sessions took part in the same places. All participants provided written informed consent delivered to their parents or guardians before the experimental session and oral assent was collected from children. All of the participating children had normal, or corrected to normal, vision, no history of neurological disorders, and, for the experimental group, normal intellectual profile (as reported in their medical records). Participants in the control group had no history of learning difficulties, dyslexia or attention deficit disorder (ADHD).

---

[1] A total of 22 children were initially recruited, but in two cases the calibration could not be properly completed; the participants did the reading task, but these sessions were not included in the study.

Demographic information is summarized in Table 1. The participants were not specifically trained in the use of read-aloud technology and none of them reported to use it on a regular basis.

| | Gender | | Age | | | | Grade (Primary school) | | |
|---|---|---|---|---|---|---|---|---|---|
| | Male | Female | Min | Max | M | SD | 3rd grade | 4th grade | 5th grade |
| **Typical readers** | 10 | 10 | 8 | 11 | 9.25 | 0.85 | 4 | 9 | 7 |
| **Dyslexic readers** | 14 | 6 | 9 | 11 | 9.55 | 0.60 | - | 13 | 7 |

[Table 1. Demographic information.]

In order to measure the representativeness of the dyslexic readers' group, the current reading profile of each participant in the experimental group was measured prior to the experiment using a standardized list of 102 Italian words and 48 Italian pseudowords (DDE-2[2];

Sartori et al., 2007). Average reading speed was 1.5 syll/sec (SD= 0.7) for words, and 1.1

syll/sec (SD= 0.3) for pseudowords; average errors were 13 (SD= 8) for words, and 12 (SD= 7)

for pseudowords. Both performance scores were below the reference values for children of the same age for all the participants in the experimental group (z-scores ranged between -0.7 and -3.2).

---

[2] DDE-2 is a widely used diagnostic test in Italy for the evaluation of developmental dyslexia and
dysorthography. In the study, we used one subtest for the evaluation of oral reading (of both words and
pseudowords).

**Material**

Participants were tested using two forms of reading material from the MT Reading Comprehension Test (Cornoldi & Colpo, 2011)[3]. This is a standardized instrument currently used in Italy for the assessment of reading processes. Its validity and reliability have been well established in the construction of the instrument and in several studies conducted by multiple investigators (e.g., Bigozzi et al., 2017; Zoccolotti et al., 2014; De Beni & Palladino, 2000). This instrument obtained good reliability (test-retest correlation above .85) and concurrent and predictive validity scores (Scorza et al., 2019). The two texts were chosen because they are part of the reference standardized test for assessing reading comprehension and performance in Italian and are targeted for fourth and fifth grade students. The two forms used in the study were similar for length (Form A: 222 words, Form B: 235 words) and complexity (Gulpease readability index for Italian text (Lucisano & Piemontese, 1988), Form A: 52; Form B: 47).

The MT Reading Test requires students to silently read a passage and then to respond to 10 multiple-choice questions that assess their understanding of the text. The answers are binary scored as correct or incorrect. Considering reference values, it is important to note that normative data for the MT reading comprehension test is based on silent individual reading without the use of a TTS tool - a different task from the one adopted in this study. For this reason, comprehension scores were measured as the number of correct answers, with a maximum score of 10.

---

[3] The two texts from the MT Comprehension Test were: Form A "Il viaggio delle anguille" ("The eels journey", 10 questions, e.g., "Which ocean do eels cross?" 222 words, GULPEASE index: 52); Form B "La croce nel cuore" ("The heart cross", 10 questions, e.g., "Where do ibex live?", 235 words, GULPEASE index: 47).

Attention-driven Read-aloud Technology

In preparing the reading material, texts from the two forms were manually segmented in groups of 1-5 words according to their prosodic phrasing, meant as the means by which speakers of any given language break up an utterance into meaningful chunks. Such phrasing, mainly driven by sentence syntax and punctuation, allows making reading expressive and fostering comprehension (Frazier et al., 2006). Following the same text prosodic phrasing, a semi-professional voice artist recorded the audio version of the texts at a reading speed of 92.5 words per minute (3.1 syl/sec) approximately.

**Procedure**

Before starting the session, the participant was welcomed by a research assistant and received a detailed description of the task and the procedure, including the information about the possibility of interrupting the experimental session at any time. After providing an oral assent to proceed, participants were specifically instructed to read the text displayed on the screen and be prepared to answer some comprehension questions. During the procedure, the child sat in a chair, approximately 50-55 cm from the screen, in a quiet room. The first activity was a 12-point calibration of the eye-tracker that took approximately 1-2 minutes. During the calibration, participants were required to follow a circle target as it displayed in 12 different positions on the screen. Before the experimental session, participants went through a practice session in which they read a sample text with GARY and answered three sample comprehension questions. In four cases (two from the experimental group and two from the control group) the calibration was not optimal, and this was observed during the training sessions. In these cases, the calibration process was repeated and then the session regularly started.

The order of the type of technology (GARY and Traditional) and the two written text varied orthogonally within each participant: half of the participants started with GARY, and half started with Traditional; if they read text A with GARY, they read text B with Traditional and vice versa.

In both conditions, seven lines of text were displayed on the screen at any given time and the read aloud text was highlighted in yellow. Participants were asked to silently read the text, and, at the end of each text, they were given the comprehension questions. Two measures were collected: accuracy in the comprehension task and reading speed. For accuracy, the number of correct responses with respect to the 10 comprehension questions was computed. For reading speed, the total time taken to read the text divided by the number of syllables of the text (in order to normalize for different text lengths) was computed.

Between the two sessions, the participants were given a break of five minutes during which they played a non-digital memory game together with a research assistant. In order to balance possible effects of the calibration task, the 12-point calibration was performed before each reading session (for both GARY and Traditional technologies). The entire procedure took approximately 60 minutes.

## Results

The accuracy results are plotted in Figure 3: when using GARY, participants with dyslexia showed higher comprehension scores than when they used the Traditional read-aloud tool. Conversely, the comprehension scores for typical readers were not modulated by the technology. The results of the linear mixed model analysis for the two dependent variables are reported in Table 2. The analysis, conducted with the *lme4* package in R (Bates, Mächler, Bolker, & Walker, 2014), showed a significant interaction between group and technology ($F_{(1,38)} = 8.4$, $p < .01$), with a significant simple main effect of the group factor ($F_{(1,38)} = 18.19$, $p < .001$), with higher reading comprehension scores for typical readers (M=6.1 ,

SD=1.7) compared to readers with dyslexia (M=8.4 , SD=1.7). No effect of technology was observed (p = 0.29). Post-hoc analyses confirmed that for participants with dyslexia scores were different depending on the type of technology used (t(38)= 2.8, p < .05 with Tukey adjustment): comprehension scores were higher when using GARY (M= 6.8, SD= 2.3) compared to Traditional (M= 5.5, SD= 2.6). No statistically significant differences emerged for typical readers (M= 8.2, SD= 1.5 with GARY, M= 8.7, SD= 1.2 with Traditional, t(38)=-1.2, p=0.57).

As a measure of effect size, we calculated the marginal $R^2$ (the variance explained by fixed factors), as well as the conditional $R^2$ (the variance explained by fixed and random factors combined) as suggested in (Nakagawa & Schielzeth, 2017). The model explains 29% of the variance in participants' comprehension scores based on group and technology (marginal $R^2$ = 0.29, conditional $R^2$ = 0.62).

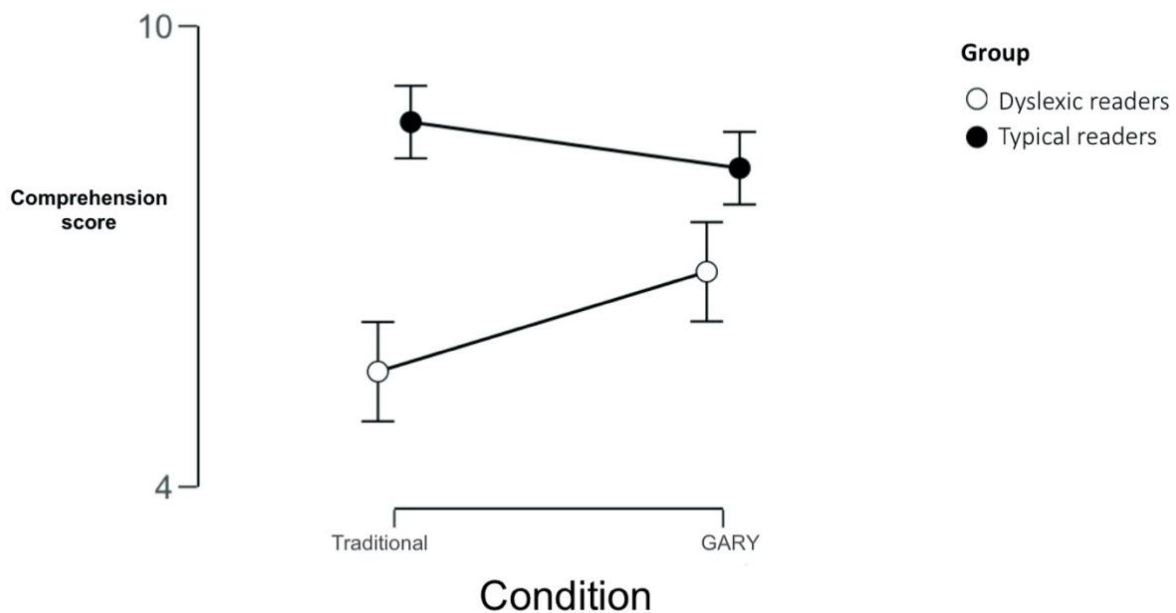

[Figure 3. Comprehension score. Error bars indicate standard errors.]

| | F | df | p |
|---|---|---|---|
| *Simple effect of Technology (GARY vs. Traditional)* | | | |
| | 1.14 | (1,38) | .292 |
| *Simple effect of Group (Typical readers vs Dyslexic readers)* | | | |
| | 18.19 | (1,38) | <.001 |
| *Interaction effect of Group x Technology* | | | |
| | 8.42 | (1,38) | .006 |

[Table 2. F values, degrees of freedom and significance levels of fixed effects in linear mixed model analyses for comprehension scores.]

For reading speed, the analyses did not show any effect of the between-group factor $(F(1,38)= 1.14, p = .29, R^2 = 0.03)$. Participants took an average of 2.3 syll/sec (SD = 0.75 syll/sec) for reading the passage with GARY. Specifically, readers with dyslexia took 2.2 syll/sec (SD= 0.66 syll/sec) and typical readers 2.4 syll/sec (SD = 0.84 syll/sec). Considering the Traditional condition, reading speed was steady at the TTS speed (i.e. 3.1 syll/sec), slightly shorter than the average speed in the GARY condition. No correlation has been observed between reading speed and comprehension scores in the GARY condition $(r_s = 0.12, p > .05)$, neither within readers with dyslexia $(r_s = -0.02, p > .05)$ or within typical readers $(r_s = 0.35, p > .05)$.

In order to better understand performance variations within the experimental group, analyses were conducted taking into account different reading profiles obtained through the performance on the standardized reading tests. Readers with dyslexia were divided into two categories based on their reading speed: High speed (N= 12) vs Low speed (N= 8), using a cut-off of fifth percentile (1.7 syll/sec); and their accuracy: High Accuracy (N= 11) vs Low Accuracy (N= 9), using a cut-off at fifth percentile (10 errors). As a dependent measure, we

Attention-driven Read-aloud Technology

calculated gain scores, i.e. difference between the comprehension scores in the GARY

condition minus the scores in the Traditional condition. On average, inaccurate readers showed

an increase in comprehension with GARY versus Traditional (M= 1.9, SD= 2.6) compared to

more accurate readers with dyslexia (M= 0.8, SD= 2.2). However, the difference is not

statistically significant (F(1,18)=1.1, p >.05). Considering the entire sample including the

control group (all typical readers are considered with high accuracy and high speed), a linear

mixed model analysis showed a significant difference in the gain scores between High (N= 31)

and Low (N= 9) Accuracy groups (F(1,38)= 6.1, p< .05, $R^2$= 0.14 - Figure 4A) but no

difference between High (N= 32) and Low (N= 8) Speed (F(1,38)= 1.6, p= 0.2, $R^2$= 0.04 -

Figure 4B). These results suggest that readers with inaccurate reading tended to benefit more by using the GARY technology compared to the Traditional condition.

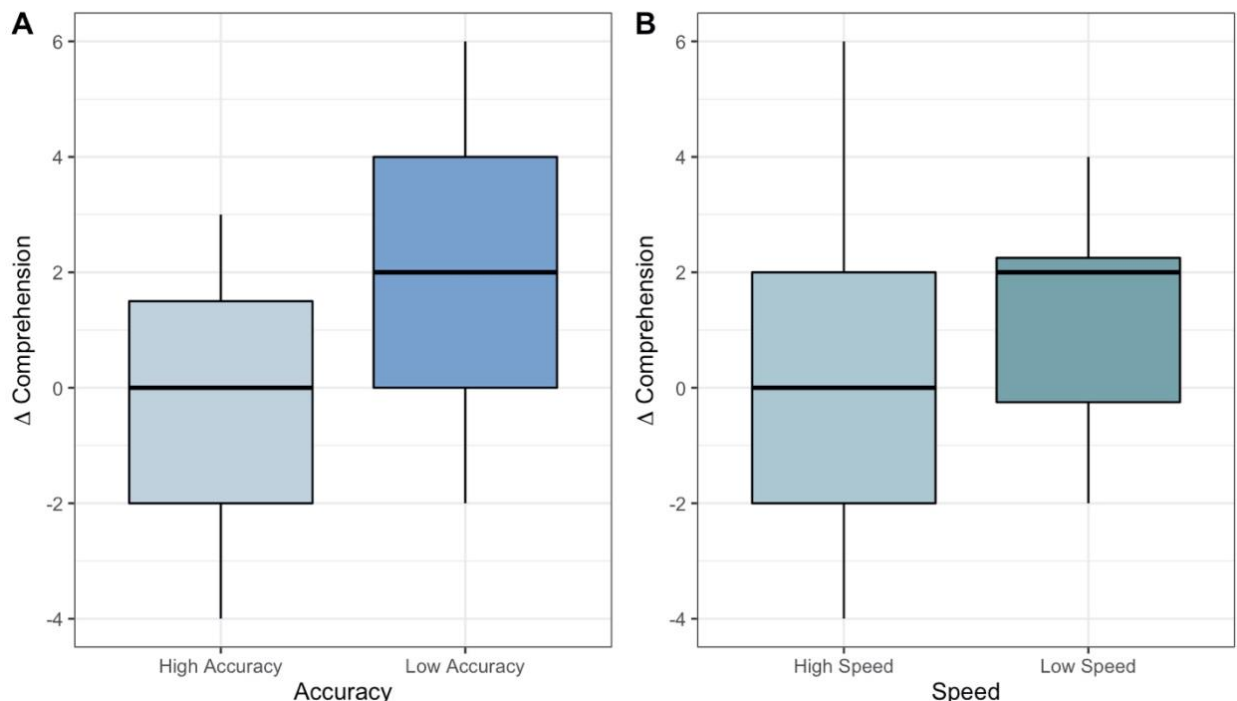

[Figure 4. Delta (change in score) between comprehension scores in GARY and Traditional.
Higher positive scores indicate better comprehension with GARY. Panel A shows High and Low
accuracy groups, panel B shows High and Low speed groups.]

During the experimental sessions, the researchers collected informal field notes during each experimental session to document how participants responded to and perceived the use of the read-aloud tools. Typical-developing readers reported a sense of unfamiliarity associated with the use of the read-aloud tools, both for the traditional TTS and GARY. This was indeed the first time they used a read-aloud accommodation tool, and its use was mostly perceived as superfluous. They also perceived GARY as slower compared to the traditional version and more difficult to use. Participants with dyslexia considered the use of read-aloud tools as not familiar either. However, they also considered reading with GARY as more adapted to their pace and "calmer" with respect to the traditional version.

## Discussion

As hypothesized, the use of GARY seems to improve reading comprehension in children with dyslexia compared to a traditional text-to-speech tool, showing an improvement of their comprehension by 24% measured on a standardized instrument for the assessment of reading comprehension.

Our main design assumption was based on the role that auditory and visual integration seems to have on reading difficulties. The fact that our approach appears to markedly benefit children with reading difficulties may suggest that this type of manipulation might be an effective intervention that is not only related to the speed of the read-aloud technology. Our approach follows the suggestion proposed by Vidyasagar (2019) of developing new remediation strategies for supporting visual attention, that is in line with recent studies on using computer games to manipulate attention and improve reading skills (Franceschini et al., 2015), and to support auditory-visual integration (Schneps et al., 2019).

Attention-driven Read-aloud Technology

Concerning dyslexic readers, we believe that our study provides initial evidence that the benefits of a read-aloud accommodation enhanced with eye-tracking are based on two main features. Firstly, by allowing readers to control the pace of the read-aloud voice by using their gaze, an attention-driven read-aloud tool might offer a more calm and effortless reading experience for struggling readers. The reading speed in the GARY condition was indeed constrained by the position of the reader's gaze and driven by his/her reading peace. This process helped the readers with dyslexia in focusing their attention on appropriate places during reading, supporting the processing of concurrent visual and auditory information, and eventually improving the reading comprehension. It should be noted that considering reading speed, the resulting reading pattern with GARY is characterized by a generally slower speed (M= 2.2 syl/sec for readers with dyslexia, M= 2.4 syl/sec for typical readers) compared to the one with the traditional text-to-speech tool (3.1 syl/sec for both groups). Struggling readers might have generally benefited from this slow and controlled modality of scanning the lines of text, while typical readers might have felt slightly distracted or annoyed by the constraints imposed by this reading modality, that did not, however, lower their performance.

Secondly, the findings seem to suggest that the use of an attention-driven read-aloud tool effectively supports the strategy of *listening-while-reading*, that is combining listening while reading (Hawkins, Marsicano, Schmitt, McCallum, & Musti-Rao, 2015; Schmitt, McCallum, Hawkins, Stephenson, & Vicencio, 2018) which is supported by recent findings that both listening to and reading a story at the same time will be maximally beneficial for word learning (Valentini et al. 2018). In using traditional read-aloud and text-to-speech applications, the reader can focus solely on the audio feedback, without the need of linking it to the visual text, helped by the dynamic highlights. In such situations, the use of the read-aloud functionality might only support comprehension from listening and not from reading. As many traditional TTS applications, GARY couples the audio feedback with the written text but it also requires the reader to explicitly gaze at the text while listening to the auditory feedback, otherwise the audio feedback and the visual highlighting will not proceed. GARY might thus support the struggling readers in visually analysing the words as they are read by the text-to-speech, helping them to decode and make the necessary connection between the sounds heard (phonemes) and the letters read (graphemes). The findings of our study show that readers with dyslexia characterized by a profile of inaccurate reading have greater grain scores in using GARY versus a traditional read-aloud tool compared to readers with dyslexia with a profile characterized by higher accuracy. This is consistent with the simple view model of reading that indicates that comprehension requires an efficient word recognition/decoding together with adequate language skills (Gough & Tunmer, 1986). Readers with dyslexia show considerable difficulties in the decoding process: the slow and inaccurate word reading can, in turn, be a bottleneck that impedes adequate reading comprehension (Snowling, 2013). The use of read-aloud tools can support the decoding process by promoting the audio-visual integration and the use of attention-driven read-aloud tools might facilitate this process by allowing readers with word decoding difficulties in combining listening and reading in an efficient and time-optimal way. Furthermore, the audio-visual integration might improve verbal and visual information processing on working memory, increasing the mental resources that can be devoted to comprehension. This might help readers to construct and maintain a situational model of the text information, connecting incoming information with the representation of the previous information in the text with the reader's own prior knowledge.

Of course, longitudinal studies are needed to evaluate and quantify the benefits of the system for a more prolonged usage. Variables such as expertise in using read-aloud tools or digital competences in general might have an effect on the efficacy of gaze-based reading support interventions, especially considering acceptance and motivation to use such tools. Other moderator factors, such as measures related to visual attention and cognitive load, should be taken into account. Moreover, we acknowledge that the system should be further assessed with respect to different measures related to reading performance, controlling for listening and complex comprehension skills.

## Conclusion

In this paper, we reported a novel approach to support children with reading difficulties by using GARY, a read-aloud technology automatically controlling the reading aloud speed coupled with the monitoring of the gaze on the text. A controlled study suggests that this attention-driven read-aloud tool improves the reading experience, in terms of performance on a standardized reading comprehension test, compared to traditional read-aloud technology, in young struggling readers. However, longitudinal studies are still needed to evaluate and quantify the benefits of the system for a more prolonged usage, controlling for the novelty effect. Moreover, the system should be further assessed with respect to different measures related to reading performance, controlling for listening and complex comprehension skills. Future studies should also take into consideration additional factors such as reading material difficulty, type of oral modality (i.e., the use of synthesized text-to-speech instead of human recorded audio), effect of reading fatigue and finally, potential moderator effect of varying levels of expertise in using text aloud tools.

Future research should further investigate the effect of using different text layouts in gaze-regulated read-aloud tools: research has shown the potential benefits of alternative modes of text presentation in digital devices, such as the Rapid Serial Visual Presentation (RSVP) approach (Koornneef et al., 2018) or the Span-Limiting Tactile Reinforcement (SLTR) intervention (Schneps, et al., 2010). In RSVP, words or full sentences of a text are displayed sequentially on a screen, often for a predetermined, limited amount of time. In SLTR the digital text is reformatted into a single column with only a few words per line, while readers are encouraged to keep their gaze at the top of the page, reading the uppermost line and then advancing the column of text using a physical button or wheel (the tactile reinforcement). RSVP approaches have been proven to benefit beginner readers (Koornneef et al., 2018), while other studies show that SLTR-based techniques improve the reading of people with dyslexia (Schenps et al., 2013). In this scenario, eye-tracking technology might be applied in synergy with RSVP and SLTR approaches to provide information on the overt attention of the reader, informing the advancement and adaptation of the serial presentation of text.

Finally, an open question remains whether the approach of gaze-regulated read-aloud could be used in intervention or in compensation setting. As observed by Wood et al. (2018), the use of text-to-speech technology can be divided into intervention and compensation studies. Intervention-oriented studies use text aloud tools to improve unassisted reading skills, whereas compensation-oriented studies address the use of text-to-speech tools to compensate for word-level skills and to help readers in accessing texts. Considering the application of attention-driven read-aloud tools, the findings from this work suggest that a gaze-regulated read-aloud technology can be used as an intervention tool for supporting students' skills in reading through text-to-speech technology. Gaze-regulated reading can be used for training readers in coupling audio and visual information and such skill could be generalized when using traditional read-aloud tools. In this respect, future studies could investigate how to use information from eye-tracking to support not only linear reading (as investigated in this study) but also strategic reading

Attention-driven Read-aloud Technology

practices (e.g., move through text, use of contextual and graphical clues, use of reference materials, etc.).

In conclusion, these findings present a promising starting point to develop adaptive applications based on eye-tracking technology for augmenting the comprehension abilities of children with reading difficulties and to improve standard read-aloud technology personalizing the reading experience.

**Acknowledgement**

The authors would like to thank Samantha Baita and Federica Maria Sole Biondi for their commitment and help in collecting the data, the Beato De Tschiderer institute and its staff for the support, and all the children who participated in the study.

Attention-driven Read-aloud Technology

Attention-driven Read-aloud Technology